\documentstyle[aps,12pt]{revtex}

\widetext
\draft
\tighten
\preprint{EFUAZ FT-99-67-REV}

\begin{document}

\title{Addendum to ``...The Relativistic Covariance of the ${\bf B}-$
Cyclic Relations" [{\it Found. Phys. Lett.} {\bf 10} (1997) 383-391]}

\author{{\bf Valeri V. Dvoeglazov}}

\address{
Escuela de F\'{\i}sica, Universidad Aut\'onoma de Zacatecas \\
Apartado Postal C-580, Zacatecas 98068, ZAC., M\'exico\\
E-mail:  valeri@ahobon.reduaz.mx\\
URL: http://ahobon.reduaz.mx/\~~valeri/valeri.htm}

\date{December 29, 1998. Revised: February 1999}

\maketitle

\begin{abstract}
In the previous paper we proved that the Evans-Vigier definitions of
$B^{(0)}$  and ${\bf B}^{(3)}$ may be related {\it not} with
magnetic fields but with a 4-vector field. In the present {\it
Addendum} it is shown that the terms used in the ${\bf B}-$ Cyclic theorem
proposed by M. Evans and J.-P. Vigier may have various transformation
properties with respect to Lorentz transformations. The fact whether the
${\bf B}^{(3)}$ field is a part of a bi-vector (which is equivalent to
antisymmetric second-rank tensor) or a part of a 4-vector, depends
on the phase factors in the definition of positive- and negative-
frequency solutions of the (${\bf B}, {\bf E}$) transverse field.
This is closely connected to our considerations of the
Bargmann-Wightman-Wigner (Gelfand-Tsetlin-Sokolik) Constructs and with the
Ahluwalia's recent consideration of the phase factor related to gravity.
The physical relevance of proposed constructs is discussed.
\end{abstract}


\newpage

In their papers and books~\cite{EV1} Evans and Vigier used
the following definition for transverse antisymmetric tensor field:
\begin{equation}
\pmatrix{{\bf B}_\perp\cr {\bf E}_\perp\cr} =
\pmatrix{{B^{(0)}\over \sqrt{2}} \pmatrix{+i\cr 1\cr 0\cr}\cr
{E^{(0)}\over \sqrt{2}} \pmatrix{1\cr -i\cr 0\cr}\cr} e^{i\phi}
+
\pmatrix{{B^{(0)}\over \sqrt{2}} \pmatrix{-i\cr 1\cr 0\cr}\cr
{E^{(0)}\over \sqrt{2}} \pmatrix{1\cr +i\cr 0\cr}\cr} e^{-i\phi}\quad,
\end{equation}
(see refs.~\cite{EV1,DVOE0} for detailed notation).
On this basis they defined so-called ${\bf B}^{(3)}$ filed
and the ${\bf B}-$ Cyclic Theorem:
\begin{equation}
{\bf B}^{(1)}\times {\bf B}^{(2)} = i B^{(0)} {\bf B}^{(3)\,\ast}
\quad et \,\, cyclic\,\, .\label{bc}
\end{equation}
This theory got great deal of criticism, see, for
instance,~\cite{Lakh0,COM1,COM2,COM3,COM4,Hunter}. Particularly,
Comay claimed~\cite{COM3} that the ${\bf B}^{(3)}$ field is
incompatible  with the Relativity Theory. I commented this
discussion in~\cite{DVOE0} and suggested that the ${\bf B}^{(3)}$ field
may be interpreted as a part of 4-vector field functions,
see also~\cite{DVOE1,DVOE2,DVOE11}.
In the present paper I show that the question of the transformation law
for such a kind of the field is not trivial and depends on the phase
factors between up- and down- parts of electromagnetic bi-vectors
(or between parts of the antisymmetric tensor which is equivalent to the
former) corresponding to the positive- and negative- frequency solutions.
{\bf The achieved result is that the} ${\bf B}^{(3)}$ {\bf field defined as
in}~(\ref{bc}) {\bf may be a part of the antisymmetric tensor field}.
In this case we encounter unusual configurations of the corresponding
transverse ${\bf B}$ and ${\bf E}$, {\it but} similar unusual
configurations of the antisymmetric tensor field have been considered for
a long time~\cite{DVOE3,KR,DVA00,DVANP,Lakh,Barr,Rodr}; they are very
well-known to the quantum-field theorists (information mainly from the
referee and editors of the  paper~\cite{DVOE11} from Physical Review D);
and a similar construct found its sound interpretation in the recent
Ahluwalia's paper~\cite{DVAF}, who is perfectly aware about previous
considerations~\cite{KR,Lakh,Barr,DVA00,DVOE3,EV1} (cf. references
in~\cite{DVAF}) and with whom we discussed all this staff during last six
years.

I would like to pass now to mathematical details.

In refs.~\cite{EV1} the authors used the  transverse solutions of
the Maxwell's equations (the formula (1) above) in order to define ${\bf
B}^{(3)}$. These tranverse solutions can be re-written to the real fields:
\begin{equation}
\pmatrix{{\bf B}_\perp\cr {\bf E}_\perp\cr} = \pmatrix{B^0 \sqrt{2}
\pmatrix{-\sin\phi\cr\cos\phi\cr 0\cr}\cr E^0 \sqrt{2}\pmatrix{\cos\phi\cr
\sin\phi\cr 0\cr} \cr} \,\, ,
\end{equation}
which represent the right-polarized
radiation ($B^0=E^0$). Of course, similar formulas exist for
left-polarized radiation.

The Lorentz transformation law for antisymmetric tensor field (written in
the form of the bi-vector) is:
\begin{equation}
\pmatrix{{\bf B}\cr {\bf E}\cr}^\prime =
\Lambda
\pmatrix{{\bf B}\cr {\bf E}\cr} =
\pmatrix{\gamma +{\gamma^2 \over
\gamma+1} \left [ ({\bf S}\cdot {\bbox\beta})^2 -{\bbox\beta}^2\right ] &
i\gamma ({\bf S}\cdot {\bbox\beta})\cr
-i\gamma ({\bf S}\cdot {\bbox\beta})&
\gamma +{\gamma^2 \over
\gamma+1} \left [ ({\bf S}\cdot {\bbox\beta})^2 -{\bbox\beta}^2\right ]
\cr}
\pmatrix{{\bf B}\cr {\bf E}\cr} \, .
\end{equation}
It is easy to see that the case considered in ref.~\cite{EV1,DVOE0}
corresponds to the choice of the field
function (operator in the quantized case) in the following form:
\begin{equation}
\pmatrix{{\bf B}_\perp\cr {\bf E}_\perp\cr}^\prime =
\Lambda \left \{ \pmatrix{\tilde{\bf B}^{(1)}\cr \tilde{\bf E}^{(1)}\cr}
e^{+i\phi} + \pmatrix{\tilde{\bf B}^{(2)}\cr \tilde{\bf E}^{(2)}\cr}
e^{-i\phi} \right \} = \Lambda \left \{ \pmatrix{\tilde{\bf B}^{(1)}\cr
-i\tilde{\bf B}^{(1)}\cr} e^{+i\phi} + \pmatrix{\tilde{\bf B}^{(2)}\cr
+i\tilde{\bf B}^{(2)}\cr} e^{-i\phi} \right \}\, .  \label{lt1}
\end{equation}
Phase factors in the formula (\ref{lt1}) is fixed between
the vector and axial-vector parts of the antisymemtric tensor field for
both positive- and negative- frequency solutions if one wants to have pure
real fields.  The ${\bf B}^{(3)}$ field in this case may be regarded as a
part of 4-vector with respect to the Lorentz transformations.

In the present {\it Addendum}
we are going to lift the above requirement and
consider the general case:
\begin{equation}
\pmatrix{{\bf B}_\perp\cr {\bf
E}_\perp\cr}^\prime = \Lambda \left \{ \pmatrix{{\bf B}^{(1)}\cr {\bf
E}^{(1)}\cr} e^{+i\phi} + \pmatrix{{\bf B}^{(2)}\cr {\bf E}^{(2)}\cr}
e^{-i\phi} \right \} = \Lambda \left \{ \pmatrix{{\bf B}^{(1)}\cr
e^{i\alpha (x^\mu)} {\bf B}^{(1)}\cr} e^{+i\phi} + \pmatrix{{\bf
B}^{(2)}\cr -e^{i\beta (x^\mu)}{\bf B}^{(2)}\cr} e^{-i\phi} \right \}\, .
\label{ltg}
\end{equation}
Our formula (6) can be re-written to the formulas generalizing
(6a) and (6b) of ref.~[2]:
\begin{mathletters}
\begin{eqnarray}
{\bf B}^{(1)\,\prime}_i &=& \left ( 1+ie^{i\alpha}\gamma ({\bf S}\cdot
{\bbox\beta}) +{\gamma^2 \over \gamma +1} ({\bf S}\cdot {\bbox \beta})^2
\right )_{ij} {\bf B}^{(1)}_j \quad,\\
{\bf B}^{(2)\,\prime}_i &=& \left ( 1 -ie^{i\beta}\gamma ({\bf S}\cdot
{\bbox\beta}) +{\gamma^2 \over \gamma +1} ({\bf S}\cdot {\bbox \beta})^2
\right )_{ij} {\bf B}^{(2)}_j \quad.
\end{eqnarray}
\end{mathletters}
Then , we repeat the procedure of ref.~[2] and find
out that the ${\bf B}^{(3)}$ field may have {\it various} transformation
laws when the transverse fields transform with the above matrix $\Lambda$.
Since the Evans-Vigier field is {\it defined} by the formula (\ref{bc})
we search the transformation law for the cross product of the transverse
modes $\left [{\bf B}^{(1)} \times {\bf B}^{(2)}\right ]^\prime =?$
with taking into account (7a,7b).
\begin{eqnarray}
\left [ {\bf B}^{(1)} \times {\bf B}^{(2)} \right ]^\prime &=&
e^{-i (\alpha -\beta)} \left [ {\bf E}^{(1)} \times {\bf E}^{(2)}\right ]
= \nonumber\\
&=&i \gamma B^{(0)} \left \{ \left [
1- {e^{i\alpha} + e^{i\beta} \over 2} (i{\bbox\beta} \cdot \hat{\bf k})
\right ] (1 + {\gamma^2 ({\bbox \beta}^2 - ({\bf S}\cdot {\bbox\beta})^2
)\over \gamma+1} )_{ij} {\bf B}^{(3)}_j +\right.\nonumber\\ &+& \left.
i{e^{i\alpha} - e^{i\beta} \over 2} ({\bf S}\cdot {\bbox\beta})_{ij} {\bf
B}^{(3)}_j - \gamma B^{(0)}\left [ i{e^{i\alpha} +e^{i\beta} \over 2} +
e^{i(\alpha+\beta)} ({\bbox\beta}\cdot \hat{\bf k})\right ]_{ij}
{\bbox\beta}_j \right \}\, .
\end{eqnarray}
We used above the definition  ${\bf B}^{(3)} = B^{(0)} \hat{\bf k}$.

One can see that we recover the formula (8) of ref.~[2] when the phase
factors are equal to $\alpha=-\pi/2$, $\beta=-\pi/2$:\footnote{
In the case
$\alpha= +\pi/2$ and $\beta=+\pi/2$, the sign of ${\bbox\beta}$
is changed  to the opposite one.}
\begin{equation}
{\bf B}^{(1)\,\prime} \times {\bf B}^{(2)\,\prime} =
{\bf E}^{(1)\,\prime} \times {\bf E}^{(2)\,\prime} =  i\gamma (B^{(0)})^2
(1- {\bbox \beta} \cdot \hat{\bf k}) \left [ \hat {\bf k} -\gamma {\bbox
\beta} +{\gamma^2  ({\bbox \beta}\cdot \hat {\bf k}) {\bbox \beta}\over
\gamma+1} \right ]\quad.
\end{equation}
But, we are able to
obtain the transformation law as for antisymmetric tensor field, for
instance when $\alpha=-\pi/2$, $\beta=+\pi/2$.\footnote{ In the case
$\alpha= +\pi/2$ and $\beta=-\pi/2$, the sign in the third term in
parentheses is changed to the opposite one.} Namely,
\begin{equation}
{\bf B}^{(1)\,\prime} \times {\bf B}^{(2)\,\prime} =
i\gamma \left [ B^{(0)}\right ]^2 \left \{
\hat{\bf k} - {\gamma {\bbox\beta} ({\bbox \beta} \cdot \hat{\bf k})\over
 \gamma + 1} + (i{\bf i} \beta_y -i {\bf j} \beta_x)\right \}\,
 \label{ltast}.
\end{equation}
The formula (\ref{ltast}) and the formula for opposite choice of phases
lead precisely to the transformation laws of
the antisymmetric tensor fields:
\begin{equation}
\left [ {\bf B}^{(3)}\right ]^\prime
= \left ( 1\pm \gamma ({\bf S}\cdot
{\bbox\beta}) +{\gamma^2 \over \gamma +1} ({\bf S}\cdot {\bbox \beta})^2
\right )_{ij} {\bf B}^{(3)}_j \quad.\\
\end{equation}
$B^{(0)}$ is a true scalar in such a case.

After D. V. Ahluwalia {\it et al.}, ref.~\cite{DVA00}, and his
commenters~\cite{DVOE3,DVOE11} (see also acknowledgements in~\cite{DVANP})
we learnt that the theory of antisymmetric tensor field 1) admits
the parity doubling; 2) suggests
various relativistic equations for its description
and 3) the third state of the field which in the massless limit can vanish
{\it only} under the certain choice of normalization and frame of
reference (the latter is valid when the instant form of relativistic
dynamics is used).

Finally, I would like to point out that the origins of such surprising
features of the antisymmetric tensor field of the second rank (unknown
until recently)  may
be 1) possible composteness of the ``photon"; 2) ``hidden"
electrodynamical non-locality (apart ref.~\cite{DVAF} see
also~\cite{Chub});  3) ``a representation space carries  more information
than a [particular] wave equation (e.g., Maxwell equations) -- as noted
also in the abstract of~\cite{DVAF}; and 4) intrinsic interlink between
gravitational and electromagnetic fields -- as noted by L. de Broglie
and G. Lochak (see~\cite{Loch}).

The question of experimental possibility of detection of the class of
antisymmetric tensor fields considered in the present {\it Addendum}
(in fact, of the {\it anti-hermitian modes} on using the terminology of
the quantum optics) is still on schedule.

In conclusion, in my opinion, all the unpleasant incidents occured during
the discussion of the ${\bf B}^{(3)}$ theory and related matters  shows
evidently serious failures of our scientific system.  Finally, I  want to
note that the topic of the ${\bf B}^{(3)}$ field (in all its
contradictions) is already well understood, in my opinion, and am not
going to enter into these discussions any more.

\acknowledgments
I am thankful to Profs. D. Ahluwalia, A. Chubykalo, E. Comay, L. Crowell,
G.  Hunter, Y. S.  Kim, other colleagues, and referees and editors of {\it
Physical Review D} for valuable discussions.  I acknowledge many internet
communications of Dr.  M. Evans (1995-96) on the concept of the ${\bf
B}^{(3)}$ field, while frequently do {\it not} agree with him in many
particular questions.  But, as shown above the postulates of the ${\bf
B}^{(3)}$ theory may be viable.

I am grateful to Zacatecas University for a professorship.
This work has been supported in part by the Mexican Sistema
Nacional de Investigadores and the Programa de Apoyo a la Carrera
Docente.

\end{document}